\documentclass[twocolumn,preprintnumbers,showpacs,amsmath,amssymb]{revtex4}
\usepackage{graphicx}
\usepackage{amssymb}

\begin{document}
\title{Quantum private comparison protocol without a third party}
\author{Guang Ping He}
\email{hegp@mail.sysu.edu.cn}
\affiliation{School of Physics, Sun Yat-sen University, Guangzhou 510275,
China}

\begin{abstract}
To evade the well-known impossibility of unconditionally secure quantum
two-party computations, previous quantum private comparison protocols have
to adopt a third party. Here we study how far we can go with two parties
only. We propose a very feasible and efficient protocol. Intriguingly,
although the average amount of information leaked cannot be made arbitrarily
small, we find that it never exceeds $14$ bits for any length of the
bit-string being compared.
\end{abstract}

\maketitle



\section{Introduction}

Private comparison, a.k.a. the socialist millionaire problem \cite{m2}, is a
two-party cryptographic task in which two millionaires Alice and Bob want to
determine if their wealth $a$ and $b$ are equal, without revealing their
actual wealth. It is analogous to a more general problem whose goal is to
compare two numbers $a$ and $b$, without revealing any extra information on
their values other than what can be inferred from the comparison result. As
a typical example of secure multi-party computations, this problem plays an
essential role in cryptography. It has many applications in e-commerce and
data mining where people need to compare numbers which are confidential.

Nevertheless, it is well-known that unconditionally secure quantum two-party
secure computations are impossible \cite{qi149,qi500,qi499,qi677,qi725,qbc14}%
, i.e., the dishonest party can always obtain a non-trivial amount of
information on the secret data of the other party. Quantum private
comparison (QPC) is also covered. To circumvent the problem, almost all
existing QPC protocols (\cite%
{qi1050,qi1043,qi1049,qi1120,qi1044,qi1045,qi1047,qi1116,qi1046,qi1048,qi1115,qi1110,qi1117,HeIJQI13,qi1192,qi1207,qi1250,qi1354}
and the references therein) added a third party to help Alice and Bob
accomplish the comparison. The only exception is Ref. \cite{qi1203}, in
which such a third party is absent. But it was later found to be insecure
\cite{HeComment15,qi1354} because the secret data of the honest party will
always be exposed completely to the other party.

While quantum cryptography has displayed promising power in many fields such
as quantum key distribution 
\cite{qi365}, it is depressing if it cannot give a good solution to this
simple problem. Also, involving a third party is undoubtfully less
convenient for the practical applications of QPC. Thus it brings us to the
question \textquotedblleft what can we achieve for QPC when it is limited to
two parties only?\textquotedblright\ Here we will propose such a protocol,
in which only a very small amount of information on the secret data being
compared will be leaked to the dishonest party. Therefore, despite that it
is not unconditionally secure, it reaches a very useful security level and
the elimination of the third party highly enhances the convenience of QPC.
As there was very little progress made during the last decade in the realm
ruled out by the existing impossibility proofs \cite%
{qi149,qi500,qi499,qi677,qi725,qbc14}, we can expect our protocol to play
more important roles in the practical applications of multi-party secure
computations.

\section{Our protocol}

Let $H(x)$\ be a classical hash function which is a $1$-to-$1$\ mapping
between the $n$-bit strings $x$ and $y=H(x)$ (i.e., $H:\{0,1\}^{n}%
\rightarrow \{0,1\}^{n}$). Denote the two orthogonal states of a qubit as $%
\left\vert 0\right\rangle _{0}$\ and $\left\vert 1\right\rangle _{0}$,
respectively, and define $\left\vert 0\right\rangle _{1}\equiv (\left\vert
0\right\rangle _{0}+\left\vert 1\right\rangle _{0})/\sqrt{2}$, $\left\vert
1\right\rangle _{1}\equiv (\left\vert 0\right\rangle _{0}-\left\vert
1\right\rangle _{0})/\sqrt{2}$. That is, the subscript $\sigma =0,1$ in $%
\left\vert \gamma \right\rangle _{\sigma }$\ stands for two incommutable
measurement bases, while\ $\gamma =0,1$\ distinguishes the two states in the
same basis. We propose the following protocol.

\bigskip

\textit{The QPC Protocol} (for comparing Alice's $n$-bit string $a\equiv
a_{1}a_{2}...a_{n}$\ and Bob's $n$-bit string $b\equiv b_{1}b_{2}...b_{n}$):

(1) Using the hash function $H(x)$, Alice calculates the $n$-bit string $%
h^{A}\equiv h_{1}^{A}h_{2}^{A}...h_{n}^{A}=H(a)$, and Bob calculates the $n$%
-bit string $h^{B}\equiv h_{1}^{B}h_{2}^{B}...h_{n}^{B}=H(b)$.

(2) From $i=1$ to $n$, Alice and Bob compare $h^{A}$ and $h^{B}$ bit-by-bit
as follows.

\qquad If $i$ is odd, then:

\qquad \qquad (2.1A) Alice randomly picks a bit $\gamma _{i}^{A}\in \{0,1\}$%
\ and sends Bob a qubit in the state $\left\vert \gamma
_{i}^{A}\right\rangle _{h_{i}^{A}}$.

\qquad \qquad (2.2A) Bob measures it in the $h_{i}^{B}$\ basis and obtains
the result $\left\vert \gamma _{i}^{B}\right\rangle _{h_{i}^{B}}$. He
announces $\gamma _{i}^{B}$\ while keeping $h_{i}^{B}$\ secret.

\qquad \qquad (2.3A) Alice announces $\gamma _{i}^{A}$.

\qquad If $i$ is even, then:

\qquad \qquad (2.1B) Bob randomly picks a bit $\gamma _{i}^{B}\in \{0,1\}$\
and sends Alice a qubit in the state $\left\vert \gamma
_{i}^{B}\right\rangle _{h_{i}^{B}}$.

\qquad \qquad (2.2B) Alice measures it in the $h_{i}^{A}$\ basis and obtains
the result $\left\vert \gamma _{i}^{A}\right\rangle _{h_{i}^{A}}$. She
announces $\gamma _{i}^{A}$\ while keeping $h_{i}^{A}$\ secret.

\qquad \qquad (2.3B) Bob announces $\gamma _{i}^{B}$.

\qquad (2.4) If $\gamma _{i}^{A}\neq \gamma _{i}^{B}$, then they conclude
that $a\neq b$, and abort the protocol immediately without comparing the
rest bits of $h^{A}$ and $h^{B}$. Otherwise they continue with the next $i$.

(3) If Alice and Bob find $\gamma _{i}^{A}=\gamma _{i}^{B}$ for all $%
i=1,...,n$ then they conclude that $a=b$.

\bigskip

\section{Correctness}

In step (2) of the protocol, when $\left\vert \gamma _{i}\right\rangle
_{h_{i}}$ is measured in the basis $h_{i}$, the results will always be $%
\gamma _{i}$. But if it is measured in a different basis $\bar{h}_{i}$, both
the results $\gamma _{i}$ and $\bar{\gamma }_{i}$ can occur with the equal
probabilities $1/2$. Therefore, when Alice and Bob input the same secret
data $a=b$, there will be $h_{i}^{A}=h_{i}^{B}$ and thus $\gamma
_{i}^{A}=\gamma _{i}^{B}$ for all $i=1,...,n$, so that the protocol will
always output the correct comparison result.
On the other hand, if $a\neq b$%
, the hash values $H(a)$ and $H(b)$ will be different too. As $H(a)$ and $%
H(b)$ distribute randomly in $\{0,1\}^{n}$,\ the case that they have $j$\ ($%
j\in \lbrack 1,n]$) different bits will occur with probability%
\begin{equation}
p_{j}=\left(
\begin{array}{c}
n \\
j%
\end{array}%
\right) /(2^{n}-1).
\end{equation}%
Consequently, in each of the corresponding $j$\ rounds of step (2), the
qubit will be prepared in the state $\left\vert \gamma _{i}\right\rangle
_{h_{i}}$ while being measured in the basis $\bar{h}_{i}$, and has
probability $1/2$ to give the result $\bar{\gamma }_{i}$, causing the protocol to
abort and output the correct comparison result $a\neq b$. The only exception
is that all these $j$\ qubits still give the result $\gamma _{i}$ though
measured in the basis $\bar{h}_{i}$. In this case the protocol will
incorrectly output the result $a=b$ while they are actually inequal. This
will occur with probability $(1/2)^{j}$. Summing over all possible $j$ values, the total probability for getting an incorrect comparison
result is%
\begin{eqnarray}
p_{inc} &=&\sum_{j=1}^{n}p_{j}\left( \frac{1}{2}\right) ^{j}  \nonumber \\
&=&\sum_{j=1}^{n}\left(
\begin{array}{c}
n \\
j%
\end{array}%
\right) \frac{1}{(2^{n}-1)2^{j}}.
\end{eqnarray}%
Therefore, our protocol is not an ideal one. Fortunately, in practical
applications $n$ is generally very high, so that $p_{inc}$ will be very
close to $0$. For example, when comparing two $32$-bit strings (i.e., $n=32$%
), the above equation shows that the incorrect comparison result occurs with
probability $p_{inc}\simeq 0.01\%$\ only.

Also, if the protocol finally outputs $a=b$ in step (3) but Alice and Bob
want to further verify this result, they can replace $a$ and $b$ with $%
a\oplus s$\ and $b\oplus s$, respectively, and run the entire protocol all
over again from step (1). Here $s$ is a random $n$-bit string chosen by
either party and announced to the other party. Then whether $a$ indeed
equals $b$ will be judged more precisely so that the probability $p_{inc}$
for finally getting a wrong comparison result will be further reduced. But
it is obvious that the price is that more information on the secret data $a$
and $b$ will be leaked as they are reused in the protocol for many times.
For simplicity, in the following we merely study the amount of information
leaked when the protocol is run only once.

\section{Security}

There is surely no security problem to consider when $a=b$, because both
parties naturally know the secret data of each other from the comparison
result. Now let us study how much information is leaked when the protocol
outputs $a\neq b$.

An important feature of our protocol is that, if it aborts after running $m$
($1\leq m\leq n$) rounds of step (2), the last $n-m$ bits of $h^{A}$ and $%
h^{B}$ will not be compared any more. They will not be encoded with qubits
and transferred in the protocol at all. Thus it is obvious that these bits
remain completely secret to the other party. As a consequence, the amount of
mutual information leaked to each party is $m$ bits at the most.

To increase $m$, a dishonest party surely wants to make the protocol abort
as late as possible. Therefore, if Alice (Bob) cheats, in each of the odd
(even) rounds of step (2.3) she (he) will always announce $\gamma
_{i}^{A}=\gamma _{i}^{B}$\ even if the actual result is $\gamma _{i}^{A}\neq
\gamma _{i}^{B}$, so that the protocol will never abort in these rounds.
In each of the rest $k\equiv m/2$ even rounds (for dishonest Alice) or
$k\equiv (m+1)/2$ odd rounds (for dishonest Bob) among the first $m$ rounds
of step (2), the cheater has to announce $\gamma _{i}$\ in step (2.2) before
the other party does so in step (2.3). Since he wishes to announce a value
that satisfies $\gamma _{i}^{A}=\gamma _{i}^{B}$\ so that the protocol will
not abort, he needs to guess the $\gamma _{i}$\ value of the qubit state $%
\left\vert \gamma _{i}\right\rangle _{h_{i}}$ that the other party sent him
in step (2.1). That is, he needs to distinguish between the density matrices
$\rho _{\gamma _{i}=0}\equiv (\left\vert 0\right\rangle
_{h_{i}=0}\left\langle 0\right\vert +\left\vert 0\right\rangle
_{h_{i}=1}\left\langle 0\right\vert )/2$\ and $\rho _{\gamma _{i}=1}\equiv
(\left\vert 1\right\rangle _{h_{i}=0}\left\langle 1\right\vert +\left\vert
1\right\rangle _{h_{i}=1}\left\langle 1\right\vert )/2$. Calculations show
that the maximal probability for distinguishing these two density matrices is%
\begin{equation}
p_{\max }=\cos ^{2}(\pi /8)\simeq 0.8536.
\end{equation}%
Therefore, the dishonest party can force each of these $k$ rounds to
continue with probability $p_{\max }$\ at the most. Consequently, the
probability for the protocol to abort at the $m$th round (i.e., it happens
to continue for the first $k-1$ even (odd) rounds while aborts at the $k$-th
even (odd) round when Alice (Bob) cheats) is%
\begin{equation}
p_{abort}^{m}=p_{\max }^{k-1}(1-p_{\max })=\cos ^{2(k-1)}(\pi /8)\sin
^{2}(\pi /8).
\end{equation}%
That is, with probability $p_{abort}^{m}$\ the protocol will abort at the $m$%
th round, and the dishonest party will know nothing about the rest $n-m$
bits of the other's hash value, so that he learns $m$ bits of information at
the most. Summing over all possible $m$ values and recall that $%
m=2k$ ($m=2k-1$) for dishonest Alice (Bob), the upper bound of the average
amount of mutual information leaked is%
\begin{eqnarray}
I_{A} &=&\sum_{k=1}^{[n/2]}m\times p_{abort}^{m}  \nonumber \\
&=&\sum_{k=1}^{[n/2]}2k\cos ^{2(k-1)}(\pi /8)\sin ^{2}(\pi /8)  \label{IA}
\end{eqnarray}%
for dishonest Alice, or%
\begin{eqnarray}
I_{B} &=&\sum_{k=1}^{[(n+1)/2]}m\times p_{abort}^{m}  \nonumber \\
&=&\sum_{k=1}^{[(n+1)/2]}(2k-1)\cos ^{2(k-1)}(\pi /8)\sin ^{2}(\pi /8)
\label{IB}
\end{eqnarray}%
for dishonest Bob. Here $[x]$ means the integer part of $x$. The above $I_{A}
$\ and $I_{B}$\ are different because our protocol is asymmetric. In the $i=1
$ round of step (2) Alice prepares the qubit, and she does not announces $%
\gamma _{i}^{A}$ until Bob announces $\gamma _{i}^{B}$. Thus dishonest Alice
can always ensure that the protocol will continue at least to the second
round. But dishonest Bob cannot completely avoid the protocol abort in the first round,
making him obtain less information than dishonest Alice.

In Fig. 1 we show $I_{A}$\ and $I_{B}$ as a function of the length $n$ of
the bit-strings $a$ and $b$ being compared. Intriguingly, the result reveals
that the average amount of mutual information leaked never excesses $14$
bits (for dishonest Alice) or $13$ bits (for dishonest Bob) for any length $n
$ of $a$ and $b$. This is because when $a\neq b$, the protocol stands a very
high probability 
to abort after merely a few rounds of step (2). Thus the rest bits of $h^{A}$
and $h^{B}$ will remain secret to the other party as their information are
never announced in the protocol.

\begin{figure}[tbp]
\includegraphics{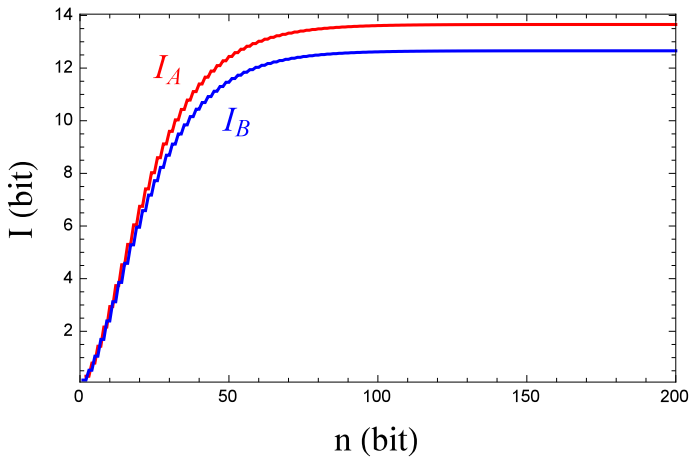}
\caption{The average amount of information leaked $I_{A}$\ ($I_{B}$) for dishonest
Alice (Bob) as a function of the length $n$ of the bit-strings that they
compare.}
\label{fig:epsart}
\end{figure}

Also, $I_{A}$\ and $I_{B}$ are merely loose upper bounds, because in the
above calculations we assumed that when the protocol aborts at the $m$th
round of step (2), the cheater obtains $m$ bits of information. But actually
this will not occur with probability $1$. For example, the optimal operation
that can distinguish the density matrices $\rho _{\gamma _{i}=0}$\ and $\rho
_{\gamma _{i}=1}$ with probability $p_{\max }=\cos ^{2}(\pi /8)$ is in fact
the measurement that projects the state $\left\vert \gamma _{i}\right\rangle
_{h_{i}}$ in the basis $\{\cos (\pi /8)\left\vert 0\right\rangle _{0}+\sin
(\pi /8)\left\vert 1\right\rangle _{0},\cos (5\pi /8)\left\vert
0\right\rangle _{0}+\sin (5\pi /8)\left\vert 1\right\rangle _{0}\}$. But
this operation provides absolutely zero knowledge on the value of $h_{i}$.
Other operations (such as the honest operation in the protocol) can learn $%
h_{i}$ with a higher probability, but result in a lower probability for
distinguishing $\rho _{\gamma _{i}=0}$\ and $\rho _{\gamma _{i}=1}$. While
the number of cheating strategies could be numerous, there is no doubt that
its probability for learning $h_{i}$ in each round can never exceed $1$, and
it can distinguish $\rho _{\gamma _{i}=0}$\ and $\rho _{\gamma _{i}=1}$ (and
thus avoid the protocol from aborting) with a probability no greater than $%
p_{\max }$. Therefore, as our above analysis assumes that the dishonest
party can distinguish $\rho _{\gamma _{i}=0}$\ and $\rho _{\gamma _{i}=1}$
with the probability $p_{\max }$ exactly while always know $h_{i}$ with
certainty, the bounds $I_{A}$\ and $I_{B}$ are sufficiently general to cover
any kind of cheating strategies.

Moreover, note that the information leaked is not the direct information on
the strings $a$ and $b$\ being compared, but the information on their hash
values instead. Thus it is less useful for the cheater in practice. For
example, even if dishonest Alice knows the exact values of the first $I_{A}$%
\ bits of Bob's $h^{B}$, it does not mean that she knows the first $I_{A}$\
bits of Bob's $b$. There are still $2^{n-I_{A}}$ bits of $h^{B}$ remain
unknown, meaning that there are $2^{n-I_{A}}$ possible choices for $b$.
Consequently, in general Alice can hardly even know whether there is $a>b$
or $a<b$, needless to say other subtle properties of $b$, such as parity,
weigh, etc.

\section{Comparison with existing QPC protocols}

As pinpointed out in Section 4 of Ref. \cite{HeIJQI13} and also in Ref. \cite%
{qi1192}, many previous QPC protocols with a third party 
also have a non-trivial amount of information leaked to at least one of
the parties. Most of them are also less feasible as they require the use of
entanglement, joint measurements, and quantum memory. On the contrary, our
protocol can easily be implemented without these resources. For example, we
can use $0%
{{}^\circ}%
$, $90%
{{}^\circ}%
$, $45%
{{}^\circ}%
$, $135%
{{}^\circ}%
$ polarized photons as the carrier of the qubit states $\left\vert
0\right\rangle _{0}$, $\left\vert 1\right\rangle _{0}$, $\left\vert
0\right\rangle _{1}$, $\left\vert 1\right\rangle _{1}$ in the protocol,
respectively. The receiver can measure them immediately, without the need to
store them. Also, in step (2) Alice and Bob send a single qubit in each
round. Once the protocol aborts, they no longer need to send qubits to
compare the rest bits of $h^{A}$\ and $h^{B}$. The fact that the average
amount of information leaked is below $14$ bits indicates that the protocol
averagely takes only $14$ rounds to abort. That is, when $a\neq b$, in
average\ our protocol merely takes $14$ qubits to complete the comparison,
thus it is extremely efficient. More importantly, all these are accomplished
without the need for the third party.

\section*{Acknowledgements}

The work was supported in part by the National Science Foundation of
China.

\end{document}